\documentclass[12pt]{JHEP3}
\pdfoutput=1

\usepackage{amsmath,epsfig}
\usepackage{amssymb,amsfonts}
\usepackage{latexsym}
\usepackage{tocvsec2}
\usepackage{subeqnarray}
\usepackage{xcolor}

\usepackage{graphicx}
\usepackage{longtable}

\relax
\def\be{\begin{equation}}
\def\ee{\end{equation}}
\def\bea{\begin{eqnarray}}
\def\eea{\end{eqnarray}}

\newcommand\fverb{\setbox\pippobox=\hbox\bgroup\verb}
\newcommand\fverbdo{\egroup\medskip\noindent%
                        \fbox{\unhbox\pippobox}\ }
\newcommand\fverbit{\egroup\item[\fbox{\unhbox\pippobox}]}

\newcommand{\bear}{\begin{eqnarray}}

\newcommand{\eear}{\end{eqnarray}}

\newcommand{\bsea}{\begin{subeqnarray}}
\newcommand{\esea}{\end{subeqnarray}}
\newbox\pippobox

\def\6{\partial}

\def\a{\alpha}

\def\z{\zeta}
\def\sq
\def\a{\alpha}

\def\hri#1#2{\href{http://arxiv.org/abs/#1}{[ArXiv:#1]#2}}

\newcommand{\ud}{\mathrm{d}}

\allowdisplaybreaks[3]

\title{Charge transport in holography with momentum dissipation}
\author{\large B. Gout\'eraux$^{a}$\\
~\\
~\\
$^a$\href{http://www.nordita.org}{Nordita}, KTH Royal Institute of Technology and Stockholm University\\
Roslagstullsbacken 23, SE-106 91 Stockholm, Sweden
\\\\
E-mail: \email{blaise@kth.se}
}

\preprint{NORDITA-2014-7}

\abstract{In this work, we examine how charge is transported in a theory where momentum is relaxed by spatially dependent, massless scalars. We analyze the possible IR phases in terms of various scaling exponents and the (ir)relevance of operators in the IR effective holographic theory with a dilaton. We compute the (finite) resistivity and encounter broad families of (in)coherent metals and insulators, characterized by universal scaling behaviour. The optical conductivity at zero temperature and low frequencies exhibits power tails which can violate scaling symmetries, due to the running of the dilaton. At low temperatures, our model captures features of random-field disorder.}


\begin{document}

\section{Introduction}

In any translation-invariant medium with a net amount of charge, applying a small electric field will result in an infinite DC conductivity, due to the fact that momentum is not relaxed and couples to the current. From the point of view of the frequency-dependent optical conductivity,  this means that its imaginary part has a pole in $1/\omega$ and hence from the Kramers-Kr\"onig relations that its real, dissipative part contains a delta function at zero frequency. In particular, there is no Drude peak at low frequencies, as the momentum relaxation rate is identically zero.

There are a number of ways to remedy this state of affairs. As investigated in previous literature, the most direct approach is to couple the charge carriers to a parametrically larger neutral bath where their momentum can relax, for instance using probe branes \cite{Karch:2007pd,Hartnoll:2009ns,cgkkm,gk} or probe fermions \cite{ProbeFermions}. Other, more involved options are to break translation invariance, either by impurities \cite{impurities,Lucas:2014zea}, by relaxing bulk diffeomorphism invariance \cite{Vegh:2013,Davison:2013jba,Blake:2013bqa}, or by turning on spatially-dependent sources \cite{Hartnoll:2012rj,Horowitz:2012ky,Blake:2013owa,Andrade:2013gsa}.

Recently, for theories where bulk diffeomorphism invariance is broken \cite{Blake:2013bqa}, a very elegant procedure was spelled out to calculate holographically the DC conductivity and was soon thereafter extended to spatially-dependent sources \cite{Blake:2013owa,Andrade:2013gsa}. 
The derivation of the formula relies on the existence of a massless mode in the spectrum of electric perturbations, which yields a radially conserved quantity at zero frequency whose boundary value gives the DC conductivity. As it is conserved through radial evolution in the bulk, it can equally well be evaluated at the horizon. This general procedure was first explained in \cite{Iqbal:2008by}. 

The formula consists of two pieces, one due to pair creation in the quantum critical sector (and already present when translation invariance is unbroken) and another, dissipative term, proportional to the net amount of charge in the system as well as to its thermal entropy. This is a similar structure to that seen in probe branes \cite{Karch:2007pd,Hartnoll:2009ns,cgkkm,gk} where in particular the close relation with the thermal entropy of the system was pointed out in \cite{cgkkm}.

The dissipative term gives the relaxation rate of the momentum, and for holographic lattices \cite{Horowitz:2012ky,Blake:2013owa} reproduces a field theory calculation in \cite{Hartnoll:2012rj}, where it was shown using the memory matrix formalism that it was related to the retarded correlator of the operator weakly breaking translation invariance.

In the AdS$_2\times\mathbf R^{p-1}$ near-horizon region of the black holes considered in \cite{Blake:2013bqa,Andrade:2013gsa}, both terms in the DC conductivity scale identically with the temperature, are constant at leading order and dictated by the ground state entropy. Therefore, to obtain more generic behaviour, the road is clear: modify the theory to obtain non-trivial scaling solutions in the IR. Such a first step was taken in \cite{Davison:2013txa} where a linear temperature dependence of the resistivity was obtained, by coupling the massive gravity sector to a neutral scalar and thus generating a specific semi-locally critical\footnote{Which means that time scales in the IR but space does not, \cite{semilocal}.} IR (also with a linear specific heat).

The main purpose of this work is to understand better how the resistivity can scale with temperature, and which critical exponents control this scaling. We will also compare our results to general expectations on dimensional grounds and previous predictions \cite{Hartnoll:2012rj,Donos:2012ra}. To allow for more general scalings, we will combine the analyses of \cite{Blake:2013bqa,Andrade:2013gsa,Davison:2013txa} with the generic IR analysis of effective holographic theories which has been pursued in \cite{cgkkm,gk,gk2012,g2013}. In this series of works, it was argued that the most generic parameterization of translation and rotation invariant extremal phases with a conserved electric flux could be achieved by specifying three scaling exponents:\footnote{With a fourth, cohesion exponent for cohesive phases \cite{gk2012,g2013}. Related work on cohesive phases also appeared in \cite{CohOther}.} a dynamical exponent $z$ measuring the anistropy between time and space; a hyperscaling violation exponent $\theta$ measuring departure from scale invariance in the metric, and resulting in an effective spatial dimensionality $d_{\theta}=p-1-\theta$ \cite{gk,sachdev}; and a conduction exponent, which measures departure from scale invariance of the electric potential and controls the scaling of the zero-temperature, low-frequency power tail of the optical conductivity. This leads to the following scaling behaviour for these fields
\be
\ud s^2=r^{\frac{2\theta}{p-1}}\left[-\frac{\ud t^2}{r^{2z}}+\frac{L^2\ud r^2+\ud\vec{x}^2}{r^2}\right],\quad A=Q r^{\zeta-z}\ud t\,,
\ee
possibly accompanied by a running scalar.
Two broad classes of solutions were exhibited, depending on whether the current dual to the gauge field is a marginally relevant or irrelevant operator in the effective holographic IR theory. In the first instance, the dynamical exponent $z$ can be adjusted freely, while the conduction exponent takes a fixed value $\zeta=-d_\theta$; in the second instance, Poincar\'e invariance is restored and $z=1$, while $\zeta$ is arbitrary.

In this work, we will generalize the setup studied in \cite{Andrade:2013gsa} to include a coupling between the massless scalars and a dilaton (a neutral scalar with an exponential potential in the IR), which allows to generate hyperscaling violation as well as modulate the dimension of the dual current. The fact that the axions \footnote{In a slight abuse of language, we will refer sometimes to the massless scalars this way, though they do not violate parity in our model.} have a spatially dependent source means momentum is dissipated, since the stress-tensor is now sourced on the right-hand side of the Ward identity. Another important technical crutch is that choosing the axions to be linear in the spatial coordinates retains homogeneity of the field equations. The analysis of possible IR phases is carried out in section \ref{section:IR}. The equations of motion are given in appendix \ref{app:A} while some technical details are relegated in appendix \ref{app:B}. We leave aside the question of finding generic finite temperature completions of the ground states we describe. However, in appendix \ref{app:C}, we do report a specific analytic AdS completion with both the axions and the dilaton turned on, with either AdS$_2\times\mathbf R^{p-1}$ or semi-locally critical ground states with $\eta=1$ (which have both an entropy and a resistivity linear in temperature), where $\eta$ is defined in \eqref{DefSL}.

Then,  in section \ref{section:resistivity}, we turn to the derivation of the finite DC conductivity in this model. An important output of this computation is the nature of charge transport. When the resistivity vanishes, the system behaves like a metal. Unless the thermal pair creation contribution to the DC conductivity is parametrically larger than the dissipative term, we expect coherent transport with a sharp Drude peak (such as were seen in \cite{Hartnoll:2012rj,Horowitz:2012ky,Davison:2013jba} for instance). From scale invariance, at low frequencies\footnote{We would like to thank S. Hartnoll for clarifications on the two formul\ae\ below.}
\be\label{ACscaling1}
\sigma(\omega,T)\sim\frac{1}{i\omega+T^{\#}F\left(\omega/T\right)}\,,\quad F(0)\sim\,constant\,,\quad F(x\gg1)\sim x^\#\,,
\ee
where the two powers $\#$ are the same and positive.
Note that this assumes that the effects of momentum relaxation are weak in the IR, i.e. that the axions are irrelevant, or marginally relevant with a weak axionic charge. Otherwise, the system is an incoherent metal, with the low temperature behaviour dominated by the quantum critical contribution from pair creation:
\be\label{ACscaling2}
\sigma(\omega,T)\sim T^{\#}G\left(\omega/T\right)\,,\quad G(0)\sim\,constant\,,\quad G(x\gg1)\sim x^\#\,,
\ee
where this time $\#<0$.
On the other hand, if the resistivity blows up at zero temperature, the system behaves like a soft-gapped insulator (earlier examples of which can be found in \cite{Donos:2012js,Donos:2013eha}), with $\#>0$ in \eqref{ACscaling2}. These last two cases are expected to correspond to strong momentum relaxation effects in the IR. We shall see whether this is borne out when the running scalar is included.

Finally, in section \ref{section:ACconductivity} we analyze the zero temperature, small frequency behaviour of the real part of the AC conductivity. Metals are expected to develop a delta function, a signal that dissipation turns off at exactly zero temperature. On top of that, a power tail exists, which a priori can come both with a positive or negative exponent. The first case is encountered for gapless, translation-invariant systems \cite{cgkkm,gk,g2013} and there all the spectral weight is transfered from the Drude peak to the delta function as the temperature is lowered. If the power tail still decays after breaking translations invariance, the scale-invariant predictions \eqref{ACscaling1}, \eqref{ACscaling2} are necessarily violated and this might hint to the reappearance of a delta function, whose origin warrants further exploration. On the other hand, when the tail blows up at low frequencies, some spectral weight remains which swamps out the delta function. It should not blow up faster than $1/\omega$ though, in order not to violate sum rules on the conductivity. For insulators, there is no delta function (the DC conductivity vanishes) and consequently the power tail should decay as well. 

We conclude in section \ref{section:ccl}, and comment on how our model captures certain features of random-field disorder at low temperatures.


\section{IR analysis for axion-dilaton theories \label{section:IR}}

Consider the following theory
\be\label{action}
S=\int\ud^{p+1}x\,\sqrt{-g}\left[R-\frac12\partial\phi^2-\frac14Z(\phi)F^2+V(\phi)-\frac12Y(\phi)\sum_{i=1}^{p-1}\partial\psi_i^2\right].
\ee
 Translation invariance is broken by the axions acquiring a (bulk) vev on-shell.
In \cite{Andrade:2013gsa}, this theory was pointed out to be not quite gauge-equivalent to massive gravity at the linear level and nonzero momentum.\footnote{This can be understood from the fact that the scalar kinetic term used in \eqref{action} only reproduces the $Tr[\mathcal K^2]$ mass term of nonlinear massive gravity and not the accompanying $Tr[\mathcal K]^2$, necessary to have a ghost-free combination. We thank A. Schmidt-May for discussions on this point.} Since we are mainly interested in zero momentum conductivities, this will not play a role in our discussion and we expect similar results would be obtained in the context of massive gravity.

We wish to look for possible IR geometries. To retain homogeneity, we will assume the axions to take the form 
\be
\psi_i=k x_i\,,\quad i=1\dots p-1
\ee
where $i$ runs over boundary spatial coordinates and $k$ can be taken identical for all $i$ without loss of generality.\footnote{Otherwise just define $k=\sqrt{\sum k_i{}^2}$.}
They correspond to marginal operators in the UV boundary CFT, with a linear source.\footnote{It would however be interesting to engineer a setup where they would be a relevant deformation while retaining homogeneity. But as we will see shortly, they can be irrelevant in the IR, just like the current.} This means that we are not describing a lattice (there is no distinguished lattice wavevector), but perhaps this model can capture features of quenched disorder at low temperatures and frequencies, like holographic massive gravity \cite{Vegh:2013,Davison:2013jba,Blake:2013bqa,Davison:2013txa}. We will come back to this interpretation in section \ref{section:ccl}.

Solutions can be distinguished along several criteria:
\begin{itemize}
\item Hyperscaling solutions where $\phi=\phi_\star$ in the IR,\footnote{We will not explicitly consider these in our analysis, since they give rise to AdS$_2\times\mathbf R^{p-1}$ in the IR, see \cite{Andrade:2013gsa}. But it should be clear how are results reduce to this case by taking the limit $z\to+\infty$ while keeping other scaling exponents finite.} or hyperscaling violating solutions where $\phi$ runs logarithmically. In that case, we approximate the scalar couplings in the IR by\footnote{All known supergravity truncations have couplings which are combinations of exponentials.}
\be
Z(\phi)\sim e^{\gamma\phi},\qquad V(\phi)\sim V_0 e^{-\delta\phi},\qquad Y(\phi)\sim e^{\lambda\phi}
\ee
and $\gamma$, $\delta$ and $\lambda$ will be related to the scaling exponents of the solutions: $z$, $\theta$ and $\zeta$.
\item (Marginally) relevant or irrelevant current, which means working out whether terms originating from the Maxwell stress-tensor in the field equations appear at the same order in powers of the radial coordinate as terms coming from the metric and neutral scalar, or are subleading. 
\item (Marginally) relevant or irrelevant axions, which means working out whether terms originating from the axion stress-tensor appear at the same order in powers of the radial coordinate as terms coming from the metric and neutral scalar, or are subleading. 
\end{itemize}

As translation invariance is not broken in the metric, the same scaling exponents as in \cite{g2013} are sufficient to describe the possible solutions, while capturing the scaling of the deformations also requires to introduce the scaling of the axion-dilaton coupling $\kappa\lambda$. They will generically take the form
\be
\ud s^2=r^{\frac{2\theta}{p-1}}\left[-f(r)\frac{\ud t^2}{r^{2z}}+\frac{L^2\ud r^2}{r^2f(r)}+\frac{\ud\vec{x}^2}{r^2}\right],\quad A=Q r^{\zeta-z}\ud t\,,\quad \phi=\kappa\ln r\,.
\ee
We relegate their precise expression in appendix \ref{app:B}. There are four classes of solutions\,.
\begin{itemize}
\item Class I, \eqref{solClassI}: both the current and the axions are (marginally) relevant in the IR. $\theta$ and $z$ are not fixed, while $\zeta=-d_\theta$ and $\kappa\lambda=-2$. This last condition is equivalent to $\gamma = (2-p) \delta +(1-p)\lambda $. It would be interesting to explore if such a condition can be understood in terms of generalized dimensional reductions \cite{gk,Gouteraux:2011qh,gk2012}.
\item Class II, \eqref{solClassII}: the current is irrelevant, the axions (marginally) relevant. $\theta$, $z$ and $\zeta\neq-d_\theta$ are not fixed, while $\kappa\lambda=-2$. This class has the remarkable property that it can display anisotropy ($z\neq1$), which is not sourced by charge density (the current is irrelevant).
\item Class III, \eqref{solClassIII}: the current is marginally relevant, the axions irrelevant. $\theta$, $z$ and $\kappa\lambda\neq-2$ are not fixed, but $\zeta=-d_\theta$.
\item Class IV, \eqref{solClassIV}: both the current and the axions are irrelevant. $\zeta\neq-d_\theta$  and $\kappa\lambda\neq-2$ are not fixed, while $z=1$.
\end{itemize}
Similarly to \cite{g2013}, we find that the conduction exponent is fixed whenever the current is (marginally) relevant. So is the axion-dilaton coupling when the axions are (marginally) relevant.

\FIGURE{
\begin{tabular}{cc}
\includegraphics[width=.45\textwidth]{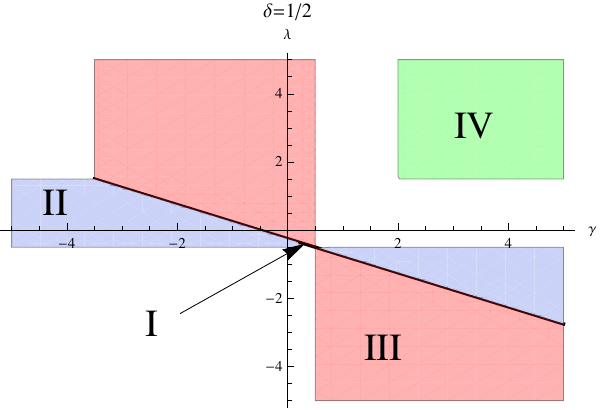} & \includegraphics[width=.45\textwidth]{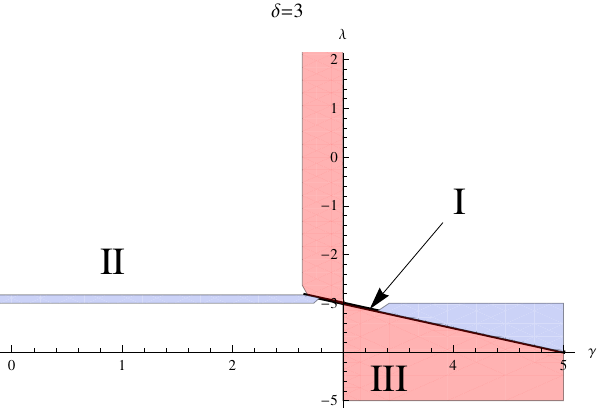}
\end{tabular}
\caption{Parameter space for classes of IR solutions, for fixed $\delta$ (left pannel: $\delta=1/2$; right pannel: $\delta=3$), in terms of $\gamma$ (horizontal axis) and $\lambda$ (vertical axis). Observe that class I appears only as a line in these plots.}
\label{fig:ParSpace}
}

In classes I and II, the axionic charge $k$ appears explicitly in the leading solution and we might expect the effects of momentum relaxation to be strong, leading to incoherent metals and insulators. In classes III and IV, the axions only appear as a deformation above the solutions of \cite{cgkkm,gk} and momentum relaxation is IR-irrelevant, so we should expect coherent metals with sharp Drude peaks.

None of these solutions compete in the same region of the parameter space ($\delta,\gamma,\lambda$), cf. figure \ref{fig:ParSpace}. We have defined the parameter space in the following way
\begin{enumerate}
\item The solution is real;
\item It has positive specific heat, which, through the scaling of entropy with temperature $S\sim T^{\frac{d_\theta}z}$, means $d_\theta/z>0$;
\item It has only irrelevant deformations, except for the temperature deformation which should be relevant.
\end{enumerate}
Within this parameter space, they all obey the NEC and the $tt$ and $x^ix^i$ elements of the metric scale the same way with $r$, so the IR is unambiguous. We can work out the spectrum of deformations along the lines of \cite{gk2012,g2013}: the conjugate modes always sum to $z+d_\theta$ as expected on dimensional grounds, with a temperature deformation associated to (marginal) time rescalings. Consequently, a blackness function can be turned on as
\be
f(r)=1-\left(\frac{r}{r_h}\right)^{z+d_\theta}\,,
\ee
when the other deformations are turned off. The parameter spaces in appendix \ref{app:B} always take into account the fact that all other deformations should be irrelevant.

Whenever $z\neq1$ (so for classes I, II and III), a semi-locally critical limit can be taken (possibly also involving $\zeta$)
\be\label{DefSL}
\theta\to+\infty\,,\quad z\to+\infty\,,\quad \frac{\theta}{z}=-\eta\,.
\ee
For classes I and II, this imposes $\lambda=0$, so a constant IR axion-dilaton coupling. In this limit, the entropy scales like $T^{\eta}$, so a linear specific heat is obtained when $\eta=1$.


\section{Resistivity \label{section:resistivity}}

\subsection{Derivation of the formula}

Let us now perturb linearly the metric and other fields by turning on a small electric field along the $x_1$ direction (which we call now $x$), at zero momentum. The only perturbations this sources are
\be
\delta A_x=a_x(r)e^{i\omega t},\quad g_{tx}=g(r)e^{i\omega t},\quad \delta\psi_1=\chi(r)e^{i\omega t}.
\ee
The independent linearized equations read, keeping in mind the Ansatz \eqref{AnsatzApp}:
\be
\begin{split}
&0=\frac{\omega ^2 a_x B}{D}+\frac{A' \left(-\frac{g C'}{C}+g'\right)}{D}+a_x' \left(-\frac{B'}{2 B}+\frac{(p-3) C'}{2 C}+\frac{D'}{2 D}+(\log Z)'\right)+a_x''\,,\\
&0=-\frac{2 i k^2 \omega  B g}{C D}+\frac{\omega ^2 B \chi }{D}+\left(-\frac{B'}{2 B}+\frac{(p-1) C'}{2 C}+\frac{D'}{2 D}+(\log Y)'\right) \chi '+\chi ''\,,\\
&0=-Za_x A'+\frac{g C'}{C}-g'-\frac{i Y D \chi '}{2 \omega }\,.
\end{split}
\ee
We can: substitute the constraint equation in the equation for $a_x$; take a derivative of the equation for $\chi$ and substitute the constraint; change variables to $\tilde \chi=C^{(p-1)/2}D^{1/2}B^{-1/2}Y\chi'/\omega$ and substitute $A'=q (BD)^{1/2}C^{-(p-1)/2}/Z$ to get the two following second-order differential equations:
\be
\begin{split}
&0=\left[ZC^{(p-3)/2}\sqrt{\frac{D}{B}}a_x'\right]'+a_x \left(\frac{e^{\gamma  \phi } \omega ^2 \sqrt{B} C^{\frac{p-3}{2} }}{\sqrt{D}}-q^2 \frac{\sqrt{BD}}{ C^{\frac{p+1}{2}}}\right)-\frac{1}{2} i q\frac{\sqrt{BD}}{ C^{\frac{p+1}{2}}} \tilde\chi\,,\\
&0=\left[Y^{-1}C^{(1-p)/2}\sqrt{\frac{D}{B}}\tilde\chi'\right]'+2 i k^2 q a_x\frac{\sqrt{BD}}{ C^{\frac{p+1}{2}}}+\left(\frac{ \omega ^2 \sqrt{B} C^{\frac{1-p}{2}}}{Y\sqrt{D}}-k^2\frac{\sqrt{BD}}{ C^{\frac{p+1}{2}}}\right) \tilde\chi\,.
\end{split}
\ee
From here on we follow closely the method set up in \cite{Blake:2013bqa,Andrade:2013gsa}, and refer to these works for more details.
The determinant of the mass matrix of the system of ODEs above is zero, so there is a massless mode. Its equation of motion reads
\be
\left[\sqrt{\frac{B}{D}}H\lambda_1'+\sqrt{\frac{B}{D}}C^{\frac{1-p}2}Y^{-1}\lambda_2\left(ZYC^{p-2}\right)'\right]'+\omega^2H\sqrt{\frac{B}{D}}\lambda_1=0\,,
\ee
where
\be
 H(r)=Z C^{\frac{p-3}{2}}-h_0 C^{\frac{1-p}2}Y^{-1}
\ee
and
\be
\begin{split}
&\lambda_1=\frac{q}{2ik^2}\frac{\left(\tilde\chi+\frac{2ik^2}{q}Z C^{p-2}a_x\right)}{C^{\frac{p-1}2}Y H},\quad \lambda_2=\frac{-q}{2ik^2}\frac{\left(\tilde\chi+\frac{2ik^2}{q}h_0 a_x\right)}{C^{\frac{p-1}2}Y H}\,.
\end{split}
\ee
From this, we deduce that the quantity
\be\label{Pi}
\Pi=\sqrt{\frac{B}{D}}H\lambda_1'+\sqrt{\frac{B}{D}}C^{\frac{1-p}2}Y^{-1}\lambda_2\left(ZYC^{p-2}\right)'
\ee
is radially conserved at zero frequency. Thus, it can be evaluated on the horizon. Following the same steps as \cite{Blake:2013bqa,Andrade:2013gsa} we find that if we define
\be\label{DCr}
\sigma_{DC}(r)=\lim_{\omega\to0}\left.\left(\frac{-\Pi}{i\omega\lambda_1}\right)\right|_r\,,
\ee
the DC conductivity is given by
\be
\sigma_{DC}=\sigma_{DC}(r\to+\infty)
\ee
if the boundary sits at infinity and provided we take $h_0=-q^2/k^2$.\footnote{On a technical level, this is so the differential equation obeyed by the massive mode $\lambda_2$ does not depend on $\lambda_1$ but just on $\lambda_1'$. Otherwise $\Pi$ does not asymptote to the DC conductivity in the zero frequency limit.} However \eqref{DCr} can be shown not to depend on $r$, and so can equally well be evaluated at the horizon. The fields satisfy ingoing boundary conditions (picking a radial gauge $D=B^{-1}=f$)
\be
\begin{split}
&a_x=(r-r_h)^{-i\omega/f'(r_h)}a_x^H\left[1+O(r-r_h)\right],\\
&\tilde\chi=(r-r_h)^{-i\omega/f'(r_h)}\tilde\chi^H\left[1+O(r-r_h)\right],
\end{split}
\ee
so that when evaluated on the horizon, the term proportional to $\lambda_2$ in the expression for $\Pi$ \eqref{Pi} drops out while the first proportional to $\lambda_1'$ will leave a non-trivial contribution. In the end, we find
\be\label{resistivity}
\sigma_{DC}=C_H^{\frac{p-3}2}Z_H+\frac{q^2}{k^2  Y_H C_H^{(p-1)/2}}\,,
\ee
where the subscript $H$ means the corresponding functions are evaluated at the horizon. This generalises the result found in \cite{Andrade:2013gsa} and is qualitatively similar to that of \cite{Blake:2013bqa}. There are two terms, each with their own interpretation: The first is due to pair creation in the background (which here is not the vacuum, but rather a quantum critical medium with a net amount of charge), and is already present in the theory without axions and momentum relaxation; The second diverges in the limit $k\to0$, highlighting the role of the axions in momentum relaxation and finite DC conductivity. So this second term is the contribution of the mechanism responsible for momentum relaxation to the conductivity. Moreover, it is inversely proportional to the thermal entropy as noted in \cite{Davison:2013txa}, where here the role of the horizon-dependent graviton mass is played by the axion-dilaton coupling $Y(\phi)$. As we comment in the discussion below, a similar relation between the resistivity and the thermal entropy also appears in the context of probe branes \cite{cgkkm}.

What are the typical behaviours one can expect at low temperatures? They fall into two broad classes: metals, for which the resistivity vanishes at zero temperature, which reflects the fact that momentum is no longer dissipated; and (soft-gapped) insulators, for which the resistivity blows up at zero temperature and the system localizes. Note that differently to \cite{Donos:2012js,Donos:2013eha}, these insulators are characterized by \emph{isotropic} gravity duals, which in particular means that lower-dimensional IR boundaries are not a necessary ingredient of holographic insulators (as in \cite{Donos:2012js}). Metals can be subdivided into two classes, those which come accompanied by a coherent Drude peak in the AC conductivity at low frequencies, for which the DC conductivity is set by the dissipative term in \eqref{resistivity} and translation invariance is weakly broken by an irrelevant operator  (like the irrelevant lattices of \cite{Hartnoll:2012rj,Horowitz:2012ky,Blake:2013owa}); and incoherent metals where there is no sharp Drude peak, or when \eqref{resistivity} is dominated by the quantum critical term and translation invariance is strongly broken. Coherent metals can thus be expected to be found in classes III and IV, incoherent metals and insulators in classes I and II.

From \eqref{resistivity}, when the system behaves like a coherent metal, we can easily derive the scattering time $\tau$ of the DC conductivity, which is given by
\be
\sigma_{DC}=Z_H C_H{}^{(p-3)/2}+\frac{\mathcal Q^2}{\mathcal E+P}\tau
\ee
where $\mathcal Q$, $\mathcal E$ and $P$ are the charge, energy and pressure density respectively. We obtain
\be\label{scattering}
\tau^{-1}=\frac{s}{4\pi}\frac{Y_H}{\mathcal E +P}\,.
\ee
Unlike for AdS$_2$, it will now display temperature dependence through the axion-dilaton coupling on the horizon, similarly to the massive gravity case \cite{Blake:2013bqa,Davison:2013txa}. It would be interesting to derive this scattering time using hydrodynamics of the axion theory, and check whether it coincides with \eqref{scattering}, along the lines of \cite{Davison:2013jba,Blake:2013bqa}.

\subsection{Low temperature behaviour of the resistivity}

Let us now examine its behaviour amongst the four classes of solutions worked out in section \ref{section:IR}. Remember that we can always turn on a small temperature in each of these solutions, which is related to the horizon radius by the scaling (which also follows by dimensional analysis) 
\be
r_h\sim T^{-\frac1z}\,.
\ee
The scaling we will obtain is then valid for temperatures low compared to the chemical potential  $T\ll\mu$.

\paragraph{Class I: Insulators and coherent metals (marginally relevant current and axion)\\}

Here, both terms in \eqref{resistivity} scale identically with the temperature, and
\be\label{resistivity1}
\rho\sim k^2 T^{\frac{2+d_\theta }{z}}\,.
\ee
Note that this recovers the result in \cite{Andrade:2013gsa} upon taking the limit $z\to\infty$, which yields an AdS$_2\times R^2$ geometry and a constant resistivity at low temperatures. On the other hand, taking the semi-locally critical limit $\theta=-\eta z$, $z\to+\infty$, we recover
\be\label{resisitivity1semilocal}
\rho\sim T^{\eta}\,,
\ee
which can be made linear by choosing $\eta=1$, as in \cite{Davison:2013txa}. If $\eta$ is kept arbitrary, the parameter space only allows for positive values, hence in this limit the system is always a metal, with a coherent Drude peak whose width and height are controlled by $k$. This is confirmed by explicit numerical calculations of the real part of the optical conductivity for AdS$_2$ solutions in \cite{Blake:2013bqa,Andrade:2013gsa}.

Coming back to finite $z$, within the parameter space discussed in section \ref{section:classI}, the scaling exponent of \eqref{resistivity1} can be both positive or negative, which means the system behaves as a metal or as an insulator, respectively. Moreover, the insulating behaviour can be seen to be tied to the vanishing/diverging of the gauge coupling in the IR being bounded, namely
\be\label{InsCriterion1}
\textrm{Insulators: }z<0\,,-2<d_\theta<0\,,\quad -2\,\frac{p-3}{p-1}<\kappa \gamma <2\quad\Leftrightarrow\quad 0<\zeta_I=-d_\theta <2
\ee
in terms of the gauge coupling or alternatively the conduction exponent. The value of the conduction exponent is not independent from $\theta$ here, since the current is marginally relevant \cite{g2013}.

For this class of solutions, the scaling of the scattering time with the temperature from \eqref{scattering} is identical to \eqref{resistivity1}, where we have used that in the low-temperature quantum critical theory, $\mathcal E$ and $P$ are constants at extremality. Consequently, this shows explicitly that whenever the system is metallic, the Drude peak sharpens up as the temperature is lowered. However, when $k$ is increased, we do expect the Drude peak to get smaller and wider, transferring spectral weight to higher frequencies.

\paragraph{Class II: insulators and incoherent metals  (marginally relevant axion, irrelevant current)\\}

 The DC conductivity \eqref{resistivity} reads at leading order in temperature
\be\label{DC2}
\sigma_{DC}=T^{(\zeta-2)/z}+\frac{q^2}{k^2}T^{-(d_\theta+2)/z}\,.
\ee
Here, the second term decays faster than the first at $T\to0$, which means that the low-temperature resistivity is dominated by pair creation in the quantum critical bath
\be
\label{resistivity2}
\rho\sim T^{\frac{2-\zeta }{z}}
\ee
set by the conduction exponent, \cite{g2013}. Note that in the class I solutions, this exponent is fixed to $\zeta_I=-d_\theta$, and replacing $\zeta$ by this value in \eqref{resistivity2}, we recover indeed the class I scaling \eqref{resistivity1}.

Within the parameter space discussed in section \ref{section:classII}, we also find that the exponent in \eqref{resistivity2} can take both positive or negative values, leading to metallic or insulating behaviour. As above, the insulating behaviour is tied to the gauge coupling being bounded from above and below
\be\label{InsCriterion2}
\textrm{Insulators:}\quad (p-3)\left(1-\frac{ \theta }{p-1}\right)<\kappa \gamma <2 (p-1)-\frac{2 (p-2) \theta }{p-1}
\ee
or similarly, in terms of the conduction exponent
\be
\textrm{Insulators:}\quad \zeta_I=-d_\theta <\zeta <2
\ee
where the lower bound is set by the value taken for the class I solutions $\zeta_I=-d_\theta>0$.

As discussed at the end of section \ref{section:classII}, one can take a semi-locally critical limit in this expression, upon which
\be
\label{resistivitySL2}
\rho\sim T^{-\tilde\zeta }
\ee
which always vanishes, hence the system is still metallic.

As the dissipative term will be parametrically smaller than the pair creation term at low temperatures, the metallic phases do not have a Drude peak but rather an  incoherent contribution, which is consistent with strong momentum IR relaxation in the IR (marginally relevant axions).

\paragraph{Class III: insulators and coherent metals (marginally relevant current, irrelevant axion)\\}

These geometries are deformations of those studied in \cite{cgkkm,gk}. The DC conductivity \eqref{resistivity} reads at leading order in temperature
\be\label{DC3}
\sigma_{DC}=T^{-(2+d_\theta)/z}+\frac{q^2}{k^2}T^{(-d_\theta +\kappa \lambda)/z}.
\ee
It is always dissipation-dominated at low temperatures, with the leading small-$T$ behaviour of the resistivity given by
\be\label{resistivity3}
\rho\sim T^{-\frac{\kappa \lambda -d_\theta}{z}}\,.
\ee
 As the momentum dissipation term dominates, we can naively expect to find no insulators but metals with a coherent Drude peak. However, the parameter space allows for both insulators or metals, i.e. the resistivity can blow up or vanish. Insulators are found when the axion-dilaton coupling and the conduction exponent are both bounded:
\be\label{InsCriterion3}
\textrm{Insulators:}\quad -2<\kappa \lambda <\zeta_I<0\,.
\ee

The metals are all expected to be coherent, since the dissipative term is parametrically larger than the pair creation at low temperatures. What is perhaps counter-intuitive is that the dissipative term can actually give rise to insulating behaviour.

\paragraph{Class IV: coherent metals (irrelevant current and axion)\\}
The DC conductivity \eqref{resistivity} reads at leading order in temperature
\be\label{DC4}
\sigma_{DC}=T^{\zeta -2}+\frac{q^2}{k^2}T^{-d_\theta+\kappa \lambda}.
\ee
It  is dissipation-dominated so that the resistivity reads at low temperatures:
\be\label{resistivity4}
\rho\sim T^{d_\theta -\kappa\lambda}
\ee
which means that its scaling is not set by the conduction exponent but by the dilaton-axion coupling.Within the parameter space \eqref{ParSpaceIV}, the resistivity vanishes, which indicates the system always behaves as a (coherent) metal.

\paragraph{Discussion\\}

In this section, we have seen how the DC conductivity could be dominated either by the pair creation term or the dissipation term. Their generic contribution is given by
\be\label{DCgen}
\sigma_{DC,pc}\sim T^{(\zeta-2)/z},\qquad \sigma_{DC,diss}\sim T^{(\kappa \lambda-d_\theta)/z}
\ee
which reduce to the correct values for each of the classes.

On physical grounds, we might expect to find coherent metallic behaviour when the two terms are of the same order, or when the dissipation term dominates. This is the case for the solutions in class III and IV, which is perhaps not suprising since the effects of momentum dissipation are irrelevant in the IR (like in \cite{Hartnoll:2012rj,Horowitz:2012ky,Blake:2013owa}). Remarkably, insulators can be found in class III in a certain range where both the conduction exponent and the axion-dilaton coupling are bounded by the other scaling exponents.

When the effects of momentum dissipation are strong, one may expect to find incoherent metals and insulators. This is partly verified by the solutions in class I, and fully in class II. In class I however, the two terms in the resistivity have the same temperature scaling, and thus they can be of the same magnitude temperature-wise and generate a sharp Drude peak for small enough $k$, similarly to what happens in \cite{Davison:2013jba,Blake:2013bqa,Andrade:2013gsa}. When $k$ increases, the peak should shrink down and broaden out, effectively transferring spectral weight to higher frequencies. In class I and II, insulators also appear whenever the conduction exponent is bounded by a certain range.

How does this compare to previous scaling arguments given to predict the behaviour of the conductivity \cite{Donos:2012ra} when momentum dissipation is relevant? The real part of the conductivity is given by the retarded current-current correlator
\be\label{CondGenEq}
\sigma_{DC,I}(T)\sim\lim_{\omega\to0}\frac1\omega\Im\left[\mathcal G^R_{\mathcal J^x\mathcal J^x}\left(\omega,T\right)\right]\sim T^{2\Delta_{\mathcal J^x}-1-(z+d_\theta)/z}
\ee
where $\Delta_{\mathcal J^x}$ is the real space dual dimension of the dual current $\mathcal J^x$ and the scaling takes into account the Fourier transform to frequency space in $d_{\theta}=p-1-\theta$ spatial dimensions. The scaling dimension of $\mathcal J^x$ is related by the current conservation equation to that of the density operator $\mathcal J^t$, which can be worked out from the mode analysis  in appendix \ref{app:B}:
\be\label{ScalDim1}
\Delta_{\mathcal J^x}=\Delta_{\mathcal J^t}+1-\frac1z\,,
\ee
where 
\be\label{ScalDim2}
\Delta_{\mathcal J^t}=\frac{d_\theta+\zeta}{2z}\,,
\ee
keeping in mind that the modes are quadratic in the irrelevant current/axion and that the above expressions are in units of frequency. From \eqref{ScalDim2}, it is clear that $\zeta$ characterizes deviation from the dimension of a conserved current in a scale invariant theory in $d_\theta$ spatial dimensions. Plugging \eqref{ScalDim1} and \eqref{ScalDim2} in \eqref{CondGenEq}, we recover the pair creation term of the DC conductivity \eqref{DCgen}. As we have already commented in the main text, for classes I and II where momentum dissipation is relevant, pair creation is always dominant and sets the scaling of the resistivity at low temperatures.

On the other hand, when translation breaking and momentum dissipation are irrelevant, \cite{Hartnoll:2012rj} predicted that the relaxation rate $\Gamma$ (and hence the contribution to the resistivity) should be given by
\be\label{DClattice}
\rho_{diss}\sim \Gamma=\frac{g^2 k_L^2}{\chi_{\vec P\vec P}}\lim_{\omega\to0}\frac1\omega\Im\left[\mathcal G^R_{\mathcal O\mathcal O}\left(\omega,T\right)\right]\sim T^{2\Delta+\frac2z-1-\frac{z+d_\theta}{z}}
\ee
where $g$ is the coupling constant of the translation-breaking deformation, $k_L$ is the lattice wavevector, $\chi_{\vec P\vec P}$ the static susceptibility of the momentum operator $\vec P$, $\mathcal O$ the operator dual to the lattice deformation and $\Delta$ its scaling dimension in real space and units of frequency. The term $+2/z$ comes from the dimension of $k^2_L$ in units of frequency, and the last term from the Fourier transform to frequency space {\em in $d_\theta$ spatial dimensions} to take into account hyperscaling violation. We do not have a lattice in this work, but we can still work out the scaling dimension of the irrelevant operator dual to the axion in the IR, for classes III and IV. From our analysis in appendix \ref{app:B}, a deformation $\psi=k x$ of the translation-invariant ground states generates a mode (at quadratic order) $1+\# k^2r^{2+\kappa\lambda}$ where $\#$ is a dimensionless number. So we can identify (in $d_\theta$ spatial dimensions)
\be
\Delta=\frac{z+d_\theta}{z}-\frac{2+\kappa\lambda}{2z}
\ee
which yields a relaxation rate consistent with $\sigma_{DC,diss}$ in \eqref{DCgen}. This confirms the prediction in \cite{Hartnoll:2012rj} (see also \cite{Blake:2013owa} for irrelevant lattice deformations).

It is also instructive to compare our results with the case of probe brane charge carriers studied in \cite{cgkkm,gk}, where the DBI action is used to model the dynamics of the charge carriers. This gives rise to a finite DC conductivity since there is a parametrically small number of charge carriers diluted in a neutral bath: this allows them to dissipate their momentum. The following expression was obtained
\be\label{DCprobe}
\sigma_{DC,DBI}=\frac{e^{-k\phi_\star}}{C_\star}\sqrt{q^2+C_\star^{p-1}Z_\star^2 e^{2k\phi_\star}}
\ee
where all quantities are evaluated at the turning point of the brane $r=r_\star$ and here $k$ labels the frame dependence of the metric as well as the origin of the neutral scalar (see \cite{cgkkm} for details). The important point to note is that \eqref{DCprobe} also displays two terms: the first is the contribution of the charge carriers to the DC conductivity, while the second is the pair creation term. The first is expected to dominate at high densities for massive carriers, while the other does for massless carriers. When the electric field on the boundary is small, the turning point $r_\star$ is well approximated by the horizon $r_h$. This means that the resistivity obtained from \eqref{DCprobe} bears a close relation to the thermal entropy, \cite{cgkkm}, just as in the formula \eqref{resistivity}. We can now compare the temperature dependence of the pair creation term with that of \eqref{resistivity}:
\be
\begin{split}
&\textrm{Relevant current:}\qquad \sigma_{DC,DBI}\sim T^{-(2+d_\theta)/z}\\
&\textrm{Irrelevant current:}\qquad \sigma_{DC,DBI}\sim T^{\zeta-2}
\end{split}
\ee
where the relevant solutions are the class III and class IV solutions without axions, studied in \cite{cgkkm,gk}. This precisely matches the scaling of \eqref{DCgen}, hinting that there is some universality behind how the pair creation contribution to the DC conductivity scales with temperature in various setups. The charge contribution in \eqref{DCprobe} is however quite different from \eqref{resistivity}, but this should not surprise us as the two terms have very different origins. We anticipate similar scalings would be found in massive gravity \cite{Blake:2013bqa} if the same IR analysis were performed.


\section{Optical conductivity \label{section:ACconductivity}}

Let us now turn our attention to the optical conductivity at nonzero frequencies. We recall the perturbation equations we obtained in the previous section
\be\label{PerEq2}
\begin{split}
&0=\left[ZC^{(p-3)/2}\sqrt{\frac{D}{B}}a_x'\right]'+a_x \left(\frac{e^{\gamma  \phi } \omega ^2 \sqrt{B} C^{\frac{p-3}{2} }}{\sqrt{D}}-q^2 \frac{\sqrt{BD}}{ C^{\frac{p+1}{2}}}\right)-\frac{1}{2} i q\frac{\sqrt{BD}}{ C^{\frac{p+1}{2}}} \tilde\chi\,,\\
&0=\left[Y^{-1}C^{(1-p)/2}\sqrt{\frac{D}{B}}\tilde\chi'\right]'+2 i k^2 q a_x\frac{\sqrt{BD}}{ C^{\frac{p+1}{2}}}+\left(\frac{ \omega ^2 \sqrt{B} C^{\frac{1-p}{2}}}{Y\sqrt{D}}-k^2\frac{\sqrt{BD}}{ C^{\frac{p+1}{2}}}\right) \tilde\chi\,.
\end{split}
\ee
We will not decouple them here for the generic case, but instead show that at zero temperature, these equations can be decoupled in the IR geometries of section \ref{section:IR}: more precisely, we are considering the region $\omega,T\ll\mu$, where $\mu$ is the chemical potential setting the scale of UV physics. Then, we can apply the matching argument of \cite{Donos:2012ra}, which relates the IR Green's functions to the UV current-current Green's function
\be\label{Im}
\Im\left[G^R_{\mathcal J^x\mathcal J^x}\left(\omega,T\right)\right]=\sum_I d^I \Im\left[\mathcal G^R_{\mathcal{O}_I\mathcal{O}_I}\left(\omega,T\right)\right],
\ee
where the index $I$ runs over all the irrelevant operators $\mathcal{O}_I$ coupling to the current $\mathcal J^x$. In our case, those operators are the current itself, and the scalar operator dual to the axion fields, as given by the two perturbations $a_x$ and $\chi$. So if we can diagonalize \eqref{PerEq2} in the IR geometries, we can work out the most relevant operator which will give the dominant contribution to the UV Green's function. This will yield the optical conductivity at zero temperature, and small frequency $\omega\ll\mu$. 

Actually, this needs only to be done explicitly for the class I solutions. For the other classes, since the coupling between the perturbations is only through a mass term, one can show that in the IR the non-diagonal mass term in each of the equations \eqref{PerEq2} is subleading and so can be neglected. Then, the two equations can be reformulated as Schr\"odinger equations
 using the change of variables
\be
a=a_x\sqrt{\tilde Z(\phi)}\,,\quad \tilde Z=C^{\frac{p-3}2}Z\,,\quad \tilde\chi=\bar\chi\sqrt{\tilde Y(\phi)}\,,\quad \tilde Y=Y C^{\frac{p-1}2}
\ee
supplemented by a radial change of coordinate
\be
\frac{\ud \rho}{\ud r}=\sqrt{\frac{B(r)}{D(r)}}
\ee
to the so-called Schr\"odinger coordinate. Inserting the scaling forms of the metric functions in terms of $\theta$ and $z$, on finds that
\be
\rho=r^{z}\,.
\ee
If we combine this with the fact that the IR is defined by the vanishing of the scale factor of the spatial part of the metric, $C(r)\sim r^{\frac{2\theta}{p-1}-2}$, and the condition for local thermodynamic stability $(p-1-\theta)z>0$, then we find that the IR in the Schr\"odinger coordinate $\rho$ is always located at $\rho\to+\infty$. The various Schr\"odinger potentials we will find will always scale like $1/\rho^2$ in the IR, and so will vanish there, indicating a gapless spectrum irrespective of the UV behaviour.

So generically we obtain a Schr\"odinger equation for a generic perturbation $\Psi_I$ with a dual operator $\mathcal O_I$
\be\label{SchrEq}
\Psi_I''(\rho)+\omega^2\Psi_I(\rho)-V_I(\rho)\Psi_I(\rho)=0\,,\quad V_I(\rho)=\frac{c_I}{\rho^2}+\cdots
\ee
where the dots denote subleading contributions to the Schr\"odinger potential in the IR. From this, we can extract the scaling of the imaginary part of the Green's function of $\Psi_I$
\be\label{ImGreenScaling}
\Im\left[\mathcal G^R_{\mathcal{O}_I\mathcal{O}_I}\left(\omega\ll\mu,T=0\right)\right]\sim \omega^{\sqrt{4c_I+1}}\,.
\ee
We then have to compare the various contributions from the different perturbations at small $\omega$ in \eqref{Im}, from which the real part of the optical conductivity reads
\be
\Re\left[\sigma\left(\omega\ll\mu,T=0\right)\right]=\frac1\omega\Im\left[G^R_{\mathcal J^x\mathcal J^x}\left(\omega\ll\mu,T=0\right)\right].
\ee

\paragraph{Class I (marginally relevant current and axion)\\}

The two equations \eqref{PerEq2} can be decoupled using the linear combinations
\be
\lambda_1=a-\frac{i q \bar\chi}{2 k^2}\,,\qquad \lambda_2= a-\frac{i \left(k^2+2 q^2\right)\bar\chi }{2 k^2 q}
\ee
and take the form of two Schr\"odinger equations \eqref{SchrEq}, from which we can extract the scalings\footnote{We have simplified an absolute value in the $\lambda_2$ scaling which always encloses a positive expression within the parameter space.}
\be\label{ACI}
\Im \left[\mathcal G^R_{\lambda_1\lambda_1}\right]\sim \omega^{\left|1-(2+d_\theta)/z\right|},\qquad \Im \left[\mathcal G^R_{\lambda_2\lambda_2}\right]\sim \omega^{\left|3+(d_\theta-2)/z\right|}\sim \omega^{3+(d_\theta-2)/z}\,.
\ee
Within the allowed parameter space, $\lambda_1$ is always the most relevant of the two in the IR, so that the optical conductivity scales like
\be\label{ACcond1}
\Re \left[\sigma\right]\sim \omega^{\left|1-(2+d_\theta )/z\right|-1}\,.
\ee
The next question is the sign of the exponent, as well as the sign of the expression within the absolute value. We find that the expression in the absolute value is positive whenever the gauge coupling is bounded\footnote{We thank A.~Donos for pointing this out to us.} by
\be\label{ABsCrit1}
\begin{split}
&-\frac{ (p-3)}{p-1}+\frac{ (p-2) }{p-1}\,z<\frac{\kappa \gamma}2 <1\,,\quad z<0\,,\quad \textrm{or}\\
&1<\frac{\kappa \gamma}2 <-\frac{ (p-3)}{p-1}+\frac{ (p-2) }{p-1}\,z\,,\quad z>2\,.
\end{split}
\ee
This range contains both insulating solutions \eqref{InsCriterion1} as well as metallic ones. Then, the scaling of the optical conductivity in \eqref{ACcond1} agrees with the scaling of the resisitivity we derived in \eqref{resistivity1}, as expected from the scaling argument in \cite{Donos:2012ra}.

There is however a region of the allowed parameter space, where the gauge coupling is not bounded and the resistivity vanishes at zero temperatures, such that the absolute value takes the opposite sign. In this case, we generically have a metal (vanishing resistivity) with a positive power tail in the optical conductivity, which always differs from the DC scaling. We will come back to this in the discussion.

In the semi-locally critical limit $\theta=-\eta z$, $\z\to+\infty$, the optical conductivity becomes
\be
\label{ACcondSL1}
\Re \left[\sigma\right]\sim \omega^{\left|1-\eta\right|-1}\,,
\ee
 which associates a $1/\omega$ power tail to the linear resistivity case $\eta=1$, in agreement with the argument of \cite{Donos:2012ra}. It is worth noting that in this limit, the resistivity vanishes at zero temperature and the system always describes a metal.

\paragraph{Class II (irrelevant current, marginally relevant axion)\\}

As mentioned above, the two perturbations $a$ and $\bar\chi$ decouple in the IR and can be shown to obey Schr\"odinger equations. From this, we derive the scalings
\be\label{ACII}
\Im\left[\mathcal G^R_{aa}\right]\sim \omega^{\left|1+(\zeta -2)/z\right|},\qquad \Im\left[\mathcal G^R_{\tilde\chi\tilde\chi}\right]\sim \omega^{\left|3+(d_\theta-2)/z\right|}\sim \omega^{3+(d_\theta-2)/z}\,.
\ee
The $\bar\chi$ scaling\footnote{We have simplified an absolute value which always encloses a positive expression within the parameter space.} is identical to the $\lambda_2$ perturbation of class I, while the $a$ scaling reduces to the $\lambda_1$ one upon taking $\zeta=-d_\theta$. The $a$ perturbation is the most IR-relevant if the conduction exponent is bounded by
\be
\textrm{Min}\left[4(1- z) ,0\right]<\zeta-\zeta_I<\textrm{Max}\left[4(1- z),0\right].
\ee
This range has to be further restricted to
\be
\textrm{Min}\left[2- z ,-d_\theta\right]<\zeta+d_\theta<\textrm{Max}\left[2- z,-d_\theta\right]
\ee
in order for the expression within the absolute value to have the right sign to match the resistivity scaling \eqref{resistivity2}. Outside of that range, the optical conductivity scaling differs from the resistivity scaling \eqref{resistivity2}. 

When the optical conductivity is given by the $a$ perturbation, then the system can behave both as a metal or as an insulator. When it is metallic, the power tail can decay or blow up towards $\omega\to0$, while it always decays for insulators.

When the optical conductivity scaling is given by the $\bar\chi$ perturbation, it reads
\be
\sigma_{\bar\chi}\sim\omega^{2+(d_\theta-2)/z}
\ee
and the exponent is always positive, so this power tail decays towards $\omega\to0$. The resisitivity \eqref{resistivity2} always vanishes, so we have a metal. This is consistent with the fact that the conduction exponent is not bounded, so the system does not localize.

\paragraph{Class III (marginally relevant current, irrelevant axion)\\}

We find the two following contributions to the imaginary part of the UV Green's function:
\be\label{ACIII}
\Im\left[\mathcal G^R_{aa}\right]\sim \omega^{\left|3+(d_\theta-2 )/z\right|}\sim \omega^{3+(d_\theta-2)/z},\qquad \Im\left[\mathcal G^R_{\tilde\chi\tilde\chi}\right]\sim \omega^{\left|1+(\kappa \lambda-d_\theta)/z\right|}\,.
\ee
Both can dominate the low-frequency behaviour. When the $a$ contribution does, the system is always metallic and the frequency-dependent power tail at zero temperature is always decaying.

When the $\tilde\chi$ contribution dominates, the system can be both metallic and insulating. The frequency-dependent power tail at zero temperature can both vanish or blow up at zero frequency in the metallic case, and it always vanishes in the insulating case. Moreover, the expression in the absolute value matches the resistivity scaling \eqref{resistivity3} when the axion-dilaton coupling is bounded
\be
\textrm{Min}(-2,d_\theta-z)<\kappa \lambda <\textrm{Max}(-2,d_\theta-z)
\ee
in terms of the conduction exponent.

\paragraph{Class IV (irrelevant current, irrelevant axion)\\}

We find the two following contributions to the imaginary part of the UV Green's function:
\be\label{ACIV}
\Im\left[\mathcal G^R_{aa}\right]\sim \omega^{\left|\zeta-1\right|}\sim \omega^{1-\zeta},\qquad \Im\left[\mathcal G^R_{\tilde\chi\tilde\chi}\right]\sim \omega^{\left|1-d_\theta +\kappa \lambda\right|}\sim \omega^{-(1-d_\theta +\kappa \lambda)}\,.
\ee
In both cases we have removed the absolute value, as allowed by the parameter space \eqref{ParSpaceIV}. 
Both can dominate the low-frequency behaviour.  The $\tilde\chi$ perturbation dominates when
\be
-\zeta+d_\theta<\kappa \lambda <-2
\ee
in terms of the conduction exponent. Irrespectively of which perturbation is the most relevant at low frequencies, the power tails always decay. Because of the sign inversion of the $\tilde\chi$ perturbation, the scaling of the optical conductivity at low frequencies can never match that of the resistivity \eqref{resistivity4}.

\paragraph{Discussion\\}

In \cite{g2013}, we found that there was a single operator in the IR, giving rise to the following scaling of the optical conductivity at zero temperature and low frequency
\be\label{ACscalingsTI}
\begin{split}
&\textrm{Relevant current:}\quad \Re\left[\sigma\left(\omega\ll\mu,T=0\right)\right]\sim\delta(\omega)+\omega^{\left|3+(d_\theta-2)/z\right|-1}\,.\\
&\textrm{Irrelevant current:}\quad\Re\left[\sigma\left(\omega\ll\mu,T=0\right)\right]\sim\delta(\omega)+\omega^{\left|1-\zeta\right|-1} \,.
\end{split}
\ee
Moreover, in the allowed parameter space, the expression within the absolute value was always positive and could be simplified.

In the setup of this paper, the analysis above shows that there is now an extra propagating mode due to the presence of the axions. The scalings of \eqref{ACscalingsTI} are most obviously compared to those of the class III and IV solutions \eqref{ACIII} and \eqref{ACIV}, from which it is clear that indeed the same mode is still present (the $a$ mode). This is because in these classes the axions are treated as irrelevant IR operators. Remarkably however, a mode with the same IR dimension is still present in class I and II, where the IR operators mix the $a$ and $\tilde\chi$ perturbations.

Generically, we find that the expression under the square root in \eqref{ImGreenScaling} is always a perfect square, hence it simplifies into an absolute value. The fact that this absolute value can change sign can be understood in the following way: the procedure we have just described amounts to taking the ratio of the normalisable over the non-normalisable piece of the IR perturbation, which then gives the imaginary part of the IR Green's function. These two pieces now typically come accompanied by a power of the radial coordinate which depends on the set of scaling exponents of the solution. Depending on their value, the two pieces can actually exchange roles, the non-normalisable piece becoming normalisable and vice-versa. This explains the absolute value, which accounts for the uncertainty over which piece is which. 

This has a dramatic consequence: Only one sign for the absolute value (the positive sign in our covention) for only one of the IR perturbations can match the resistivity scalings of the previous section. It turns out that the scaling of the optical conductivity at zero temperature and low frequencies can differ from the scaling of the resistivity, either because the absolute value has the wrong sign, or because the 'wrong' perturbation is the most relevant. Ultimately, this can be traced back to the presence of the running scalar and to the violation of scaling symmetries \eqref{ACscaling1}, \eqref{ACscaling2}. This is confirmed by the fact that the scaling symmetries are violated when the gauge or axion-dilaton couplings are unbounded by other scaling exponents, so that the dilaton running is 'strong'.

At the transition point where the expression in the absolute value changes sign and scale invariance \eqref{ACscaling1}, \eqref{ACscaling2} is violated, the resistivity is automatically linear in temperature with a $1/\omega$ tail in the optical conductivity, which is reminescent of the mechanism pointed out in \cite{Donos:2012ra}.

We do however find consistent behaviours. Whenever the system is insulating (so there is no delta function at zero temperature and zero frequency), the diverging of the resistivity at small temperatures is matched with a vanishing of the optical conductivity at zero temperature and small frequencies. On the other hand, when the system is metallic and the resistivity vanishes at low temperatures, one may expect the Drude peak to sharpen into a delta function at exactly zero frequency, that is
\be
\Re\left[\sigma\left(\omega\ll\mu,T=0\right)\right]\sim \delta(\omega)+\omega^{|n|-1}.
\ee
However, $|n|-1$ can actually be positive or negative. If it is positive, we have a vanishing power law and a diverging DC conductivity, so all the low energy spectral weight is transferred to the delta function.
If it is negative, there is a diverging power tail which washes out the delta function and signals that some spectral weight does remain at non zero energies. It would be interesting to verify this by numerical computations, in particular whether a delta function is indeed still present when there is a negative power tail, which cannot be inferred from the analytical calculations above (but could be on inspection of the imaginary part of the conductivity from the presence of a $1/\omega$ pole).

 
\section{Conclusion and outlook\label{section:ccl}}

In this work, we have examined how momentum can be relaxed in holographic theories containing axions with a source linear in one of the spatial coordinates. By aligning each axion along a different spatial direction, homogeneity and isotropy of the system is retained, which means that the framework set up in \cite{g2013} for the analysis of translation-invariant phases still applies.

Doing so, we have performed an analysis of the possible phases with hyperscaling violation (which naturally encompasses hyperscaling cases) and showed how it could be split up in four classes of solutions, depending on whether the current and the axions are (marginally) relevant operators in the IR or not. Each solution is captured by a set of four scaling exponents: the dynamical exponent $z$, the hyperscaling violation exponent $\theta$ and the conduction exponent $\zeta$ introduced in \cite{g2013}; as well as the axion-dilaton coupling. If the axions are marginally relevant, the axion-dilaton coupling is fixed $\kappa\lambda=-2$, while if the current is marginally relevant, it is the conduction exponent $\zeta=\theta-p+1=-d_\theta$.

Since momentum is relaxed, the theory gives rise to a finite DC conductivity and hence, resistivity. We have derived a generic formula \eqref{resistivity}, which generalizes that of \cite{Andrade:2013gsa} and is qualitatively similar to previous results in holographic massive gravity \cite{Blake:2013bqa} or probe charge carriers \cite{cgkkm,gk}. It contains two terms, one from pair creation in the vacuum and another dissipative term proportional to the charge density, which generically scale as
\be\label{DCgenccl}
\sigma_{DC,pc}\sim T^{(\zeta-2)/z},\qquad \sigma_{DC,diss}\sim T^{(\kappa \lambda-d_\theta )/z}.
\ee
Along these lines, the conduction exponent $\zeta$ should then really be thought as controlling the quantum critical, pair creation contribution to the DC conductivity. 
This term turns out to have the same scaling as that obtained from probe branes  in \cite{cgkkm}, pointing to some universality.

 If the resistivity vanishes at low temperatures, the system behaves like a metal: without a coherent Drude peak if the first term is parametrically larger than the second or for large enough axionic charge, with a coherent Drude peak otherwise. If the resistivity diverges at low temperatures, we find soft-gapped insulators, which have a translation-invariant metric and no anisotropy contrarily to those of \cite{Donos:2012js,Donos:2013eha}.

Turning to the optical conductivity, its scaling at low frequencies and zero temperatures can be determined. For insulating phases, we always find a decaying power tail. For metals however, we either find a superposition of a delta function and a decaying power tail, indicating that all the spectral weight is transferred to the delta function; or a diverging power tail broadening out the delta function. It would be interesting to work out (numerically) the frequency-dependence of the optical conductivity in more detail.

Intriguingly, these power tails do not necessarily agree with the resistivity scaling and can violate scale invariant expectations \eqref{ACscaling1}-\eqref{ACscaling2}, contrarily to hyperscaling cases \cite{Hartnoll:2012rj,Donos:2012ra,Donos:2012js}: this is a side effect of a strong running of the dilaton, happening in regions of the parameter space where the exponents $\zeta$ and $\kappa\lambda$ (or alternatively, the gauge- and axion-dilaton couplings) governing the AC conductivity are unbounded. The scaling violation can manifest itself in two ways. The first is that there are generically two modes propagating in the IR and contributing to the UV retarded Green's function of the current. Only one of the two can possibly match the resistivity scaling but either can be the most relevant depending on the parameter space. Even when the correct mode is the most relevant, its contribution to the conductivity scales like $\omega^{|n|-1}$, where it only agrees with the resistivity for $n>0$, which again is not necessarily guaranteed by the parameter space. This reflects the fact that in the IR region, the 'source' and the 'vev' (i.e. the non-normalisable and normalisable pieces) can be exchanged depending on the values of the scaling exponents. However, when the violation of scale invariance is realised in this way, the resistivity becomes linear in temperature at this transition point $n=0$, which is reminescent of the mechanism described in \cite{Donos:2012ra} (albeit at zero momentum).

One interesting consequence of our analysis is the following: whenever the resistivity is linear (like for instance in the semi-locally critical case analogous to \cite{Davison:2013txa}), the power tail in the optical conductivity goes like $1/\omega$ (as was also pointed out in \cite{Donos:2012ra}). It would be desirable to understand what consequences this has on the calculation of the sum rule on the real part of the conductivity, which is not integrable at $\omega=0$, and whether these two features can be decoupled.

In this work, we have considered spatially-dependent but linear sources for the axions, which are marginal deformations of the UV CFT. It would be very interesting to make these deformations relevant while retaining homogeneity, like in \cite{Donos:2013eha}, and investigate how and if these results change, particularly the various scaling behaviours. This also opens the way for an analysis of phases which are spatially anisotropic. Another interesting setup could involve helical (Bianchi VII) symmetries, \cite{Donos:2012js}. There, some extra scaling exponents are needed to parameterize the spatial anisotropy, but the conductivity displays similar scaling properties, in particular with negative, frequency-dependent power tails as well as insulating behaviour, \cite{BianchiVII}. Of course, efficient, power-law momentum relaxation will not always occur in this setup: lattices at finite $z$ relaxing momentum at the lattice scale should result in Boltzmann-suppressed resistivities \cite{Hartnoll:2012rj, semilocal}.

In \cite{Davison:2013txa}, it was shown that for a specific $\eta=1$ semi-locally critical geometry in holographic massive gravity, the resistivity scaled linearly like the entropy $\rho\sim s$: the explanation put forward was that if the late-time behaviour of the system is controlled by hydrodynamics,\footnote{assuming a hydrodynamic state can form, which means Umklapp scattering should occur on much longer timescales.} the momentum relaxation rate associated to quenched disorder is set by the shear viscosity, which is famously related to the entropy density via a universal ratio. Hence \cite{Davison:2013txa} concluded that massive gravity captures leading effects of quenched disorder, while subleading corrections in $1/\log T$ have to be worked out for instance using the memory matrix formalism. 
When it dominates, the dissipative term in \eqref{DCgenccl} generates a leading contribution to the resistivity
\be\label{rhodissccl}
\rho\sim s \,T^{-\frac{\kappa\lambda}{z}}\,,\qquad s\sim T^{\frac{d_\theta}z}\,.
\ee
Obviously, it reproduces the result of \cite{Davison:2013txa} in the semi-locally critical limit $z\to+\infty$.
In the presence of spatially-dependent axions, the universality of the shear viscosity to entropy ratio will be violated, but will only generate subleading corrections \cite{DiscDavison} so we can expect the previous result to still hold. It would be interesting to understand if there is some universality behind the temperature prefactor.

At finite $z$, \cite{Lucas:2014zea} analyzed the effects of random-field disorder on a generic hyperscaling violating but scale invariant theory. They found that for relevant disorder with UV scaling dimension $\Delta$, the contribution to the resistivity was $\rho\sim T^{2(1+\Delta-z)/z}$, which upon saturation of the Harris criterion (meaning that disorder becomes marginally relevant and perturbation theory breaks down), turned into $\rho\sim s\, T^{2/z}$. Remarkably, this scaling coincides with \eqref{rhodissccl} when the axions are marginally relevant in the IR and $\kappa\lambda=-2$ (classes I and II), bringing further evidence that the axions capture some of the IR physics associated with random-field disorder and can relax momentum efficiently at finite $z$.  We have worked at zero momentum throughout this paper: indeed the disorder calculations \cite{Davison:2013txa,Lucas:2014zea} are also dominated by low momenta modes. \cite{Hartnoll:2014cua} has shown that (UV) marginal disorder gave rise to Lifshitz IR geometries with $z$ finite: our results for class II solutions, which have IR marginally relevant axions and finite $z$ backgrounds, seem to resonate with the interpretation that our massless scalars capture random disorder physics at low temperatures.


\paragraph{Note added:} \cite{Donos:2014uba} appeared simultaneously where a subset of class II solutions as well as anisotropic phases analogous to class I are discussed in four bulk dimensions. The formula for the DC conductivity is also obtained for $D=4$ as well as qualitatively similar results for the scaling of the conductivity.


\acknowledgments

It is my pleasure to gratefully acknowledge insightful correspondence and discussions with M.~Blake, A.~Donos, S.~Hartnoll, E.~Kiritsis, A.~Krikun, A.~Lucas, S.~Sachdev, K.~Schalm, A.~Schmidt-May, D.~Tong, D.~van der Marel, B.~Withers and especially R.~Davison. I am particularly thankful to E.~Kiritsis and B.~Withers for their comments on a previous version of this manuscript, and to the anonymous referees for their suggestions.\\
Note v4: I thank Marco Caldarelli, Richard Davison and Li Li for pointing out typos in previous versions of the manuscript.


\appendix

\addtocontents{toc}{\protect\setcounter{tocdepth}{1}}

\section{Equations of motion\label{app:A}}

The equations of motion derived from the action
\be
S=\int\ud^{p+1}x\,\sqrt{-g}\left[R-\frac12\partial\phi^2-\frac14Z(\phi)F^2+V(\phi)-\frac12Y(\phi)\sum_{i=1}^{p-1}\partial\psi_i^2\right]
\ee
read
\be
\begin{split}
&R_{\mu\nu}=\frac12\partial_\mu\phi\partial_\nu\phi+\frac{Y(\phi)}2\sum_{i=1}^{p-1}\partial_\mu\psi_i\partial_\nu\psi_i + \frac{Z(\phi)}2F_\mu{}^\rho F_{\nu\rho}-\frac{Z(\phi)F^2}{4(p-1)}g_{\mu\nu}-\frac{V(\phi)}{p-1}g_{\mu\nu}\,,\\
&0=\nabla_\mu\left(Z(\phi)F^{\mu\nu}\right),\\
&0=\nabla_\mu\left(Y(\phi)\nabla^\mu\psi_i\right),\qquad i=1\dots p-1\,,\\
&0=\Box\phi+V'(\phi)-\frac14 Z'(\phi)F^2-\frac12Y'(\phi)\sum_{i=1}^{p-1}\partial(\psi_i)^2.
\end{split}
\ee
Plugging in the Ansatz
\be\label{AnsatzApp}
\ud s^2=-D(r)\ud t^2+B(r)\ud r^2+C(r)\ud\vec{x}^2,\quad \phi=\phi(r)\,,\quad A=A(r)\ud t\,,\quad \psi_i=k x_i\,,
\ee
the equations of motion are
\be
\begin{split}
&0=\frac{2 B V}{p-1}+\frac{Z (p-2) A'^2}{(p-1) D}+\frac{B' D'}{2 B D}-\frac{(p-1) C' D'}{2 C D}+\frac{D'^2}{2 D^2}-\frac{D''}{D}\,,\\
&0=\left[\frac{Z C^{\frac{1}{2} (p-1)}}{\sqrt{BD}}A'\right]'\,,\\
&0=\frac{B' D'}{2 B D}-\frac{(p-1) C'^2}{2 C^2}-\frac{(p-1) C' D'}{2 C D}-\frac{B' }{2B}\left(\frac{(p-1) C'}{ C}+\frac{D'}{ D}\right)+\phi '^2+\frac{(p-1) C''}{C}\,,\\
&0=\frac{Y\, k^2 B}{C}-\frac{2 B V}{p-1}+\frac{Z A'^2}{(p-1) D}+\frac{(p-3) C'^2}{2 C^2}+\frac{C'}{2C} \left(\frac{D'}{D}-\frac{B'}{B}\right)+\frac{C''}{C}\,,\\
&0=-\frac{Y_{,\phi}\, k^2 (p-1)   B}{2 C}+\frac{Z_{,\phi}  A'^2}{2 D}+B V_{,\phi}-\frac{B' \phi '}{2 B}+\frac{(p-1) C' \phi '}{2 C}+\frac{D' \phi '}{2 D}+\phi ''\,,
\end{split}
\ee
where we have suppressed the dependence of all functions on $r$ or $\phi$ for brevity, primes denote derivatives wrt $r$ and the axion equations are automatically satisfied by our Ansatz.

\section{Details on the IR analysis\label{app:B}}


\subsection{Class I: marginally relevant current, marginally relevant axion \label{section:classI}}

We start by considering that both the current and the axions are marginally relevant in the IR.
It is easy to find a scaling solution of the form
\be\label{solClassI}
\begin{split}
&\ud s^2=r^{\frac{2\theta}{p-1}}\left[-\frac{\ud t^2}{r^{2z}}+\frac{L^2\ud r^2+\ud\vec{x}^2}{r^2}\right],\quad L^2=\frac{2 (p-2+z-\theta ) (p-1+z-\theta )}{2 V_0-k^2 (p-2)}\,,\\
&A=\sqrt{\frac{2 \left(2V_0(1-z)+k^2 (p z-z-\theta) \right)}{\left(k^2( p-2)-2 V_0\right) (p-1+z-\theta )}}r^{1-p-z+\theta}\ud t\,,\\
&e^\phi=r^\kappa\,,\qquad \kappa^2=\frac{2 (p-1-\theta ) (1+p (z-1)-z-\theta )}{p-1}\,,\\
&\gamma = (2-p) \delta +(1-p)\lambda \,,\qquad \kappa\delta=\frac{2 \theta }{p-1}\,,\qquad \kappa\lambda=-2\,,
\end{split}
\ee
which differs from the scaling solutions when translation invariance is not broken \cite{cgkkm,gk}.
If one desires, a blackness function can be turned on exactly and is written
\be
f(r)=1-\left(\frac{r}{r_h}\right)^{p-1+z-\theta}\,.
\ee

Let us now run the usual mode analysis. What we get are conjugate modes, summing to $p-1+z-\theta=d_\theta+z$. Two pairs are degenerate, with a zero mode and a universal, temperature mode equal to $p-1+z-\theta$ (as expected). The last pair is more interesting and reads
\be
\begin{split}
&\beta_\pm=\frac{z+d_\theta}2\pm\sqrt{\frac{-1+p+z-\theta }{4 (1-p-z+p z-\theta )}\left(X-4\frac{ k^2}{V_1} (p-2) d_\theta  (d_\theta-1+z)\right)}\\
&X=9 p^2 (z-1)-17-9 z^2-8 \theta +\theta ^2+z (26+8 \theta )+p \left(26+9 z^2+8 \theta -z (35+9 \theta )\right)\\
&V_1= -k^2 (-2+p)/2+V_0\,.
\end{split}
\ee
$\beta_+$ will have the same sign as the temperature deformation and so is always relevant, but it is however possible to check that $\beta_-$ is always irrelevant given the parameter space defined by: real solution, relevant temperature deformation and positive specific heat.

Moreover, one can check that the $tt$ element of the metric scales like the spatial directions, that is blows up or vanishes when they do. The Null Energy Condition is always satisfied.

The allowed parameter space is
\be\label{ParSpaceI}
\begin{split}
&V_0>0\,,\quad k^2V_0<\frac{-2+2 z}{-z+p z-\theta }\,,\quad \textrm{and}\\
&z<0\,,\quad \theta >p-1\quad \textrm{or }\quad 1<z\leq 2\,,\quad \theta <(z-1)(p-1)\quad\textrm{or }\quad z>2\,,\quad\theta <p-1
\end{split}
\ee
with a maximum value for $k^2$.

\paragraph{Semi-locally critical limit\\}
In the limit
\be
\theta\to+\infty\,,\quad z\to+\infty\,,\quad \frac{\theta}{z}=-\eta
\ee
the solution \eqref{solClassI} becomes conformal to AdS$_2\times\mathbf R^{p-1}$:
\be\label{solClassISL}
\begin{split}
&\ud s^2=r^{-\frac{2\eta}{p-1}}\left[\frac{L^2\ud r^2-\ud t^2}{r^2}+\ud\vec{x}^2\right],\qquad L^2=\frac{2 (1+\eta )^2}{2 V_0-k^2 (p-2)}\,,\\
&A=\sqrt{\frac{2 \left(2V_0+k^2 (p -1+\eta) \right)}{\left(k^2( p-2)-2 V_0\right) (1+\eta )}}r^{-1-\eta}\ud t\,,\\
&e^\phi=r^\kappa\,,\qquad \kappa^2=\frac{2\eta (p-1+\eta )}{p-1}\,,\\
&\gamma = (2-p) \delta\,,\qquad \lambda=0 \,,\qquad \kappa\delta=\frac{-2 \eta }{p-1}\,,
\end{split}
\ee
with blackness function
\be
f(r)=1-\left(\frac{r}{r_h}\right)^{1+\eta}\,.
\ee
The allowed parameter space is
\be\label{ParSpaceISL}
V_0>0\,,\quad k^2V_0<\frac{2}{-1+p+\eta }\,,\quad \eta>0
\ee


\subsection{Class II: irrelevant current, marginally relevant axion \label{section:classII}}

An irrelevant current means that it backreacts as a mode on the background solution. As a consequence, the background is a solution of the equations of motion with the gauge field turned off: it will be turned on at linear order in deformations, and backreact at quadratic order on the other fields. The background in that case is a hyperscaling violating solution, characterized by a set of three scaling exponents: $z$, $\theta$ and $\zeta$, the conduction exponent. It reads
\be\label{solClassII}
\begin{split}
&\ud s^2=r^{\frac{2\theta}{p-1}}\left[-\frac{\ud t^2}{r^{2z}}+\frac{L^2\ud r^2+\tilde L^2\ud\vec{x}^2}{r^2}\right],\qquad L^2=\frac{ ((p-1)z-\theta ) (p-1+z-\theta )}{V_0}\,,\\
&e^\phi=r^\kappa\,,\qquad \tilde L^2=\frac{k^2((p-1) z-\theta)}{2 (z-1)V_0}\,,\\
& \kappa\delta=\frac{2 \theta }{p-1}\,,\qquad \kappa\lambda=-2\,,\qquad \kappa^2=\frac{2 (p-1-\theta ) ((p-1) (z-1)-\theta )}{p-1}\,.
\end{split}
\ee
A blackness function can be turned on and reads as previously
\be
f(r)=1-\left(\frac{r}{r_h}\right)^{p-1+z-\theta}\,.
\ee

Note that we can still engineer $z\neq1$, i.e. IR violation of relativistic symmetry, with an irrelevant current. This is different from the backgrounds studied in \cite{cgkkm, gk, gk2012,g2013}, where non-relativistic IR backgrounds could only be obtained through a marginally relevant current.

The mode analysis reveals pairs of conjugate modes summing to $p-1+z-\theta$. One pair is simply the marginal mode and its conjugate temperature mode. Another pair does not involve the gauge field and reads
\be
\begin{split}
&\beta_\pm=\frac{p-1+z-\theta}2\pm\sqrt{\frac{X}{4 (1+p (-1+z)-z-\theta )^2}}\,,\\
&X=8 (z-1) (1+p (z-1)-z-\theta ) \left((p-1) z^2+\theta  (1-p+\theta )+z \left(1+p^2-p (2+\theta )\right)\right)\\
&\qquad+\left(p^2 (z-1)-1+2 z-z^2+\theta ^2+p \left(2+z^2-z (3+\theta )\right)\right)^2.
\end{split}
\ee
$\beta_+$ is always relevant, $\beta_-$ irrelevant. Turning to the gauge field modes, they can be parameterized as 
\be
A=Q r^{\beta_a^\pm},\qquad \beta_a^-=0,\qquad \beta_a^-=\zeta-z\,,\qquad \kappa\gamma=p-1-\zeta -\frac{p-3}{p-1}\theta\,. 
\ee
The first is a zero mode which is just a reflection of the global U(1) inside the gauge symmetry. The second generates a constant electric flux proportional to $Q$. These modes backreact on the other fields (metric and $\phi$) at quadratic order, which allows to determine the dual dimension of the current as
\be
\beta_-=\frac12(p-1+2 z-\zeta -\theta) +\beta^a_-=\frac12(p-1+\zeta -\theta)\,.
\ee
The conjugate to $\beta_-$ is absent when the flux is conserved, simply because a constant shift in the gauge field does not backreact on the other fields. If the flux was not conserved, we could work out what $\beta_+$ is and find that it sums to the correct value with $\beta_-$, $\beta_++\beta_-=p-1+z-\theta$.

The allowed parameter space is
\be\label{ParSpaceII}
\begin{split}
&V_0>0\,,\quad \theta \leq 0\,,\quad z>1\,,\quad\zeta <1-p+\theta\quad \textrm{or }\\
&0<\theta <-1+p\,,\quad z>\frac{-1+p+\theta }{-1+p}\,,\quad\zeta <1-p+\theta\quad \textrm{or }\\
&\theta >-1+p\,,\quad z<0\,,\quad\zeta >1-p+\theta\,.
\end{split}
\ee
Once all these constraints are taken into account, the Null Energy Condition holds and the $tt$ element of the metric scales together with the spatial elements.

\paragraph{Semi-locally critical limit\\}

A semi-locally critical limit can be taken as well, upon which the axion-dilaton coupling goes to a constant in the IR $\lambda=0$. However, here the limit should also include the conduction exponent $\zeta=\tilde\zeta z$, $z\to+\infty$, in order to allow for full generality in the scaling of the electric potential.

The allowed parameter space is
\be\label{ParSpaceIISL}
V_0>0\,,\quad \eta>0\,,\quad \tilde\zeta<-\eta\,.
\ee


\subsection{Class III: marginally relevant current, irrelevant axion \label{section:classIII}}

Let us now consider the following possibility: the current is marginally relevant, but the axion is not. This generates a background characterized by $z$ and $\theta$, with a mode turning on the axion:
\be\label{solClassIII}
\begin{split}
&\ud s^2=r^{\frac{2\theta}{p-1}}\left[-\frac{\ud t^2}{r^{2z}}+\frac{L^2\ud r^2+\ud\vec{x}^2}{r^2}\right],\qquad L^2=\frac{ (p-2+z-\theta ) (p-1+z-\theta )}{2 V_0}\,,\\
&A=\sqrt{\frac{2 (-1+z)}{-1+p+z-\theta }}r^{1-p-z+\theta}\ud t\,,\\
&e^\phi=r^\kappa\,,\qquad \kappa^2=\frac{2 (p-1-\theta ) (1+p (z-1)-z-\theta )}{p-1}\,,\\
&\kappa\gamma =2 (p-1)-\frac{2 (p-2) }{p-1} \theta \,,\qquad \kappa\delta=\frac{2 \theta }{p-1}\,.
\end{split}
\ee
This background is exactly identical to those discussed in \cite{cgkkm,gk}, where the axion fields where not turned on. This is because in this case they behave as deformations.

The same remarks as before regarding the blackness function and the semi-locally critical limit apply. Turning to the mode analysis, we find three pairs of modes summing to $p-1+z-\theta$: two are degenerate, with a marginal mode and a temperature mode equal to $p-1+z-\theta$. Another pair reads
\be
\begin{split}
&\beta_\pm=\frac12(p-1+z-\theta)\pm\sqrt{\frac{X}{4(2 p (-1+z)-2 (-1+z+\theta ))}}\,,\\
&X=8 (p-1) (z-1)  \left(2+p^2+z^2+p (2 z-3-2 \theta )+3 \theta +\theta ^2-z (3+2 \theta )\right)\\
&\qquad+\frac{\left(-1+p^2 (-1+z)+2 z-z^2+\theta ^2+p \left(2+z^2-z (3+\theta )\right)\right)^2}{(1+p (z-1)-z-\theta )}\,,
\end{split}
\ee
where always one among $\beta_+$ or $\beta_-$ is irrelevant, depending on the region of the parameter space determined by: real solution, relevant temperature deformation and positive specific heat. Moreover, the Null Energy Condition always holds, and the $tt$ and spatial metric elements always scale together.

Turning to the axion, it generates a mode 
\be
\beta=\kappa\left(\frac{\gamma +(p-2)\delta }{p-1}+\lambda\right)=2+\kappa\lambda
\ee
which becomes marginal precisely when the axion cannot be considered as irrelevant, and yields the solutions found previously. However, there are non-trivial constraints on the value of $\lambda$, it is not always consistent to deform the geometries of \cite{cgkkm,gk} by axions not coupled to the dilaton. The deformation can be relevant and lead to the geometries discussed above.

The parameter space implies $V_0>0$ as well as:
\be\label{ParSpaceIII}
\begin{split}
z<0\,,\qquad\theta >-1+p\,,\qquad\kappa \lambda >-2&\quad\textrm{ or}\\
1<z\leq 2\,,\qquad\theta <1-p-z+p z\,,\qquad\kappa \lambda <-2&\quad\textrm{ or}\\
z>2\,,\qquad\theta <-1+p\,,\qquad\kappa \lambda <-2\,.&
\end{split}
\ee


\subsection{Class IV: irrelevant current, irrelevant axion \label{section:classIV}}

We finally turn to the last possibility, which is that both the current and the axion are irrelevant. Then we find a hyperscaling violating solution with $z=1$
\be\label{solClassIV}
\begin{split}
&\ud s^2=r^{\frac{2\theta}{p-1}-2}\left[-\ud t^2+L^2\ud r^2+\ud\vec{x}^2\right],\qquad L^2=\frac{ (p-1-\theta ) (p-\theta )}{V_0}\,,\\
&e^\phi=r^\kappa\,,\qquad\kappa\delta=\frac{2 \theta }{p-1}\,,\qquad \kappa^2=\frac{2\theta (1+\theta-p )}{p-1}
\end{split}
\ee
It has two degenerate pairs of conjugate modes, one marginal and the other a temperature mode, which sum to $p-\theta$. Then, there are two gauge field modes
\be
\beta_-^a=0\,,\qquad \beta_+^a=\zeta-1\,,\qquad \kappa\gamma=p-1-\zeta -\frac{p-3 }{p-1} \theta
\ee
which backreact on the metric (for the non-constant mode as):
\be
\beta_-=p-1+\zeta -\theta\,.
\ee
Finally, the axion mode reads
\be
\beta=2+\kappa\lambda\,.
\ee
Note that it would impose a non-trivial constraint on the location of the IR if $\lambda=0$. The consistent parameter space is simple
\be\label{ParSpaceIV}
\theta<0\,,\quad \zeta<\theta+1-p\,,\quad \kappa\lambda<-2
\ee
and of course since $z=1$, the $tt$ element of the metric always scales in concert with the spatial ones. The Null Energy Condition always holds.


\section{Analytic asymptotically AdS family\label{app:C}}

Consider the following theory
\be
S=\int\ud^{p+1}x\,\sqrt{-g}\left[R-\frac12\partial\phi^2-\frac14Z(\phi)F^2+V(\phi)\right].
\ee
When
\be
Z(\phi)=e^{-(p-2)\delta\phi}\,,\quad V(\phi)=V_1e^{\frac{\left((p-2)(p-1) \delta ^2-2\right) \phi}{2 (p-1) \delta }} +V_2e^{\frac{2 \phi }{\delta -p \delta }}+V_3e^{(p-2) \delta  \phi }\,,
\ee
with
\be
\begin{split}
&V_1=\frac{8 (p-2) (p-1)^2 V_0 \delta ^2}{p \left(2+(p-2)(p-1) \delta ^2\right)^2}\,,\quad V_2=\frac{ (p-2)^2 (p-1) V_0 \delta ^2 \left(p(p-1) \delta ^2-2\right)}{p \left(2+(p-2)(p-1) \delta ^2\right)^2}\,,\\
&V_3=-\frac{2  V_0 \left((p-2)^2 (p-1) \delta ^2-2 p\right)}{p \left(2+(p-2)(p-1)\delta ^2\right)^2}\,,
\end{split}
\ee
then there is an analytic black hole solution, \cite{gk,Hendi:2010gq} (setting $V_0=p(p-1)$ from now on)
\be
\begin{split}
&\ud s^2=-f(r)h(r)^{\frac{-4}{2+(p-2)(p-1) \delta ^2}}\ud t^2+h(r)^{\frac{4}{(p-2) \left(2+(p-2)(p-1)\delta ^2\right)}}\left[\frac{\ud r^2}{f(r)}+r^2\ud\Sigma^2_{\kappa,p-1}\right],\\
&f(r)=r^2 \left(h(r)^{\frac{4 (p-1)}{(p-2) \left(2+(p-2)(p-1)\delta ^2\right)}}-\left(\frac{r_h}r\right)^{p} h(r_h)^{\frac{4 (p-1)}{(p-2) \left(2+(p-2)(p-1) \delta ^2\right)}}\right)+\kappa \left(1-\left(\frac{r_h}r\right)^{p-2}\right),\\
&e^\phi=h(r)^{\frac{-2(p-1)\delta }{2+(p-2)(p-1)\delta ^2}}\,,\qquad h(r)=1+\frac{Q}{r^{p-2}}\,,\\
&A(r)=2\sqrt{\frac{(p-1)Q}{p-2}}\frac{\sqrt{r_h^{2+p} h(r_h)^{\frac{2 \left(2-(p-2)^2 (p-1) \delta ^2\right)}{(p-2) \left(2+(p-1)(p-2) \delta ^2\right)}}+r_h^{p} \kappa h(r_h)^{-1}}}{r_h^{p-1} h(r)\sqrt{2+(p-2)(p-1) \delta ^2}}\left(1-\frac{r_h^{p-2}}{r^{p-2}}\right).
\end{split}
\ee
Here the horizon can be flat $\kappa=0$, positively or negatively curved.

We are interested in generalising the above, for the flat case, to include axions aligned along horizon directions. The metric (with $\kappa=0$) looks the same, while the functions which are modified read:
\be
\begin{split}
&f(r)=r^2 \left(h(r)^{\frac{4 (p-1)}{(p-2) \left(2+(p-1)(p-2) \delta ^2\right)}}-\frac{r_h^p}{r^{p}} h(r_h)^{\frac{4 (p-1)}{(p-2) \left(2+(p-1)(p-2) \delta ^2\right)}}\right)+\frac{k^2 \left(1-\frac{r_h^{p-2}}{r^{p-2} }\right)}{2(p-2)}\,,\\
&A(r)=2\sqrt{(p-1)Q}\frac{\sqrt{(p-2)r_h^{2+p} h(r_h)^{\frac{2 \left(2-(p-2)^2 (p-1) \delta ^2\right)}{(p-2) \left(2+(p-1)(p-2) \delta ^2\right)}}- \frac{r_h^{p}k^2}{2 h(r_h)}}}{(p-2)r_h^{p-1} h(r)\sqrt{2+(p-2)(p-1) \delta ^2}}\left(1-\frac{r_h^{p-2}}{r^{p-2}}\right),\\
&\psi_i=kx^i\,,
\end{split}
\ee
where $x^i$ are the boundary spatial directions. The similarity between the role of the axionic charge and the horizon curvature in the two metrics is striking. This family of solutions with the dilaton turned off was studied recently in \cite{Andrade:2013gsa}, and earlier in \cite{Bardoux:2012aw} (see also \cite{ConfCoupled} for families of axion-dilaton solutions with non-minimal couplings between the gravity and dilaton sectors).

The chemical potential can be read off from the asymptotic value of the electric potential:
\be
\mu=2\sqrt{(p-1)Q}\frac{\sqrt{(p-2)r_h^{2+p} h(r_h)^{\frac{2 \left(2-(p-2)^2 (p-1) \delta ^2\right)}{(p-2) \left(2+(p-1)(p-2) \delta ^2\right)}}- \frac{r_h^{p}k^2}{2 h(r_h)}}}{(p-2)r_h^{p-1}\sqrt{2+(p-2)(p-1) \delta ^2}}
\ee
and defines a maximum value for $k$ at fixed $Q$ and $r_h$:
\be
k^2_{max}=2 (p-2)h(r_h)^{-1+\frac{4 (p-1)}{(p-2) \left(2+(p-2) (p-1) \delta ^2\right)}}r_h^2\,.
\ee
This is similar to what we have seen in the class I solutions in section \ref{section:IR}.

The temperature reads:
\be
4\pi T=\frac{h(r_h)^{\frac{-2 (p-1)}{(p-2) \left(2+(p-2) (p-1) \delta ^2\right)}}}{r_h}\left|\frac{k^2}2-\frac{r_h^2\left(4 (p-1)+h(r_h) (p-2) \left(p(p-1) \delta ^2-2\right)\right)}{\left(2+(p-2) (p-1) \delta ^2\right)h(r_h)^{1-\frac{4 (p-1)}{(p-2) \left(2+(p-2) (p-1) \delta ^2\right)}}}\right|.
\ee

Depending on the value of $\delta$, one may check that the near-horizon geometry can be either AdS$_2\times R^2$,  or conformal to AdS$_2$ for $\delta=\sqrt{2/p(p-1)}$. In this case, it has $\eta=1$ and displays both a linear resistivity and a linear entropy in temperature.

It would be very interesting to search for other analytic AdS completions, perhaps along the lines of \cite{Anabalon:2013sra}.


\end{document}